\documentclass[conference]{IEEEtran}

\IEEEoverridecommandlockouts

\usepackage{amsmath,amsthm,relsize}
\usepackage{amsfonts, amssymb, cuted}
\usepackage{cleveref}
\usepackage{mathtools}
\usepackage{algorithm}
\usepackage[noend]{algpseudocode}
\usepackage{cite}
\usepackage{xcolor}
\usepackage{soul}
\usepackage{graphicx}
\usepackage{pgfplotstable}
\usepackage{pgfplots}
\usepackage{nicefrac}
\usepackage{enumitem}
\usepackage{support-caption}
\usepackage[style=base,font=small,skip=1pt,belowskip=-8pt]{caption}
\usepackage{multirow}
\pgfplotsset{compat=1.14}
\newcommand{\subparagraph}{}
\usepackage{titlesec}
\titlespacing\section{3pt}{6pt plus 4pt minus 2pt}{6pt plus 2pt minus 2pt}
\titlespacing\subsection{3pt}{4pt plus 4pt minus 2pt}{4pt plus 2pt minus 2pt}
\titlespacing\subsubsection{3pt}{3pt plus 4pt minus 2pt}{0pt plus 2pt minus 3pt}

\theoremstyle{definition}

\newcommand{\CA}[0]{{\mathcal{A}}}
\newcommand{\CB}[0]{{\mathcal{B}}}

\newcommand{\CD}[0]{{\mathcal{D}}}

\newcommand{\CF}[0]{{\mathcal{F}}}

\newcommand{\CJ}[0]{{\mathcal{J}}}

\newcommand{\CL}[0]{{\mathcal{L}}}

\newcommand{\CN}[0]{{\mathcal{N}}}

\newcommand{\CS}[0]{{\mathcal{S}}}
\newcommand{\CT}[0]{{\mathcal{T}}}

\newcommand{\CZ}[0]{{\mathcal{Z}}}

\newcommand{\Bh}[0]{{\mathbf{h}}}

\newcommand{\Bu}[0]{{\mathbf{u}}}
\newcommand{\Bv}[0]{{\mathbf{v}}}
\newcommand{\Bw}[0]{{\mathbf{w}}}
\newcommand{\Bx}[0]{{\mathbf{x}}}

\newcommand{\BV}[0]{{\mathbf{V}}}

\begin{document}

\title{Subpacketization-Beamformer Interaction\\ in Multi-Antenna Coded Caching
\thanks{This work was supported by the Academy of Finland under grants no. 319059 (Coded Collaborative Caching for Wireless Energy Efficiency) and 318927 (6Genesis Flagship).}}
\date{November 2019}
\author{
\IEEEauthorblockN{
MohammadJavad Salehi\IEEEauthorrefmark{1},
Antti T\"olli\IEEEauthorrefmark{1},
Seyed Pooya Shariatpanahi\IEEEauthorrefmark{2}
}\\ \vspace{-4mm}
\IEEEauthorblockA{\IEEEauthorrefmark{1}Center for Wireless Communications, University of Oulu, Oulu, Finland.}\\ \vspace{-4mm}
\IEEEauthorblockA{\IEEEauthorrefmark{2}School of Electrical and Computer Engineering, College of Engineering, University of Tehran, Tehran, Iran.}\\ \vspace{-4mm}
\IEEEauthorblockA{\{fist\_name.last\_name\}@oulu.fi; p.shariatpanahi@ut.ac.ir}
}
\maketitle

\begin{abstract}
We study the joint effect of beamformer structure and subpacketization value on the achievable rate of cache-enabled multi-antenna communications at low-SNR. A mathematical approach with low-SNR approximations is used, to show that using simplistic beamformer structures, increasing subpacketization degrades the achievable rate; in contrast to what has been shown in the literature for more complex, optimized beamformer structures. The results suggest that for improving the low-SNR rate, subpacketization and beamformer complexity should be jointly increased.
\end{abstract}

\begin{IEEEkeywords}
Coded Caching,
Multi-Antenna Communications,
Subpacketization,
Beamformer Design
\end{IEEEkeywords}

\section{Introduction}
Network data volume has continuously grown during the past years. The global IP (Internet Protocol) data volume is expected to exceed 4.8 Zettabytes ($10^{21}$ bytes) by 2022, from which 71 percent will pass through wireless networks~\cite{cisco2018cisco}. On the other hand, introduction of new application types for 5G and beyond (e.g. autonomous vehicles, immersive viewing and massive machine-type communications) has necessitated extreme advancements for all networking KPIs (Key Performance Indicators) such as data rate, delay and reliability \cite{6Genesis2019KeyIntelligence}. This has imposed serious challenges in all data network layers, and solving them is one of the main recent research trends.

Coded Caching (CC), recently proposed in \cite{maddah2014fundamental}, is considered as a solution to higher data rate requirements, specially for the prominent use-case of video-based applications. The idea 
%behind CC 
is to transmit carefully designed codewords over shared data links; enabling an additional global caching gain, proportional to the total cache size in the network, to be achieved in addition to the local caching gain at each node. Interestingly, it is shown that CC gain is additive with the multi-antenna gain \cite{shariatpanahi2018physical}, making it even more desirable for next-generation wireless networks.

Coded caching has been studied extensively in the literature. A major part of the studies is dedicated to the information-theoretic analysis; i.e. finding the maximum possible CC gain under various assumptions \cite{karamchandani2016hierarchical,pedarsani2016online}, as well as designing less complex solutions for achieving a near-optimal performance \cite{lampiris2018adding,salehi2019coded}. Specifically, reducing subpacketization, defined as the number of smaller parts each file should be split into, is considered for both single- and multi-antenna CC setups; and is shown to be nicely achievable in cache-enabled multi-antenna networks \cite{lampiris2018adding,salehi2019coded}. However, these studies consider error-free, same-capacity communication links for all users.

From another perspective, CC performance at low-SNR is also addressed in the literature. In \cite{tolli2018multicast} it is shown that the achievable rate of CC setups at low-SNR can be improved considerably, by using optimized beamformers instead of zero-forcing (ZF). In \cite{salehi2019subpacketization} a flexible-subpacketization CC design is introduced and it is shown that the achievable rate is improved, as subpacketization is increased. The results are only valid for the case of optimized beamformers though.

In this paper we extend the results of \cite{salehi2019subpacketization}, by studying the joint effect of subpacketization value and beamformer structure on low-SNR performance. Interestingly, we show that the positive effect of increased subpacketization might be completely reversed, if simpler beamformers are used instead of the optimized ones. We also mathematically analyze the observations; using proper low-SNR approximations. Our results provide a better understanding of various parameters affecting the performance of CC setups in real-world implementations.

Throughout the text, we use $[K]$ to denote  $\{1,2,...,K\}$ and $[i:j]$ to represent $\{i,i+1,...,j\}$. Boldface upper- and lower-case letters denote matrices and vectors, respectively. $\BV[i,j]$ refers to the element at the $i$-th row and $j$-th column of matrix~$\BV$. Sets are denoted by calligraphic letters. For two sets $\CA$ and $\CB$, $\CA \backslash \CB$ is the set of elements in $\CA$ which are not in $\CB$; and $|\CA|$ represents the number of elements in $\CA$.

\section{System Model and Literature Review}
\subsection{Problem Setup}

We consider a multiple input, single output (MISO) setup. A server, equipped with $L$ transmitting antennas, communicates with $K$ single-antenna users over a shared wireless link. The server has access to the file library $\CF$, which has $N$ files each with size $f$ bits. Every user $k \in [K]$ has a cache memory of size $Mf$ bits, where $M \le N$. For simplicity, we use a normalized data unit and drop $f$ in subsequent notations.

The system operation consists of two phases; placement and delivery. At the placement phase, 
%which takes place at the low network traffic time,
cache memories of the users are filled with data from the files in $\CF$.
This is done without any knowledge of file request probabilities in the future; and hence an efficient strategy is to store equal-sized data portions of all files (with size $\nicefrac{M}{N}$) in the cache memory of each user.
%. This is done in accordance with a cache placement algorithm; and
%without any knowledge of file request probabilities in the delivery phase. 
We use $\CZ(k)$ to denote the cache contents of user $k$.
%, after the placement phase is completed.

At the beginning of the delivery phase, every user $k$ reveals its requested file $W(k) \in \CF$. 
%Let us 
Consider
%define the demand set as 
$\CD = \{W(k) \mid k \in [K] \}$
to be the demand set.
Based on $\CD$ and $\CZ(k)$,
%and in accordance with a delivery algorithm
the server builds and transmits (e.g. in a TDMA fashion) transmission vectors $\Bx(i) \in \mathbb{C}^L, i \in [I]$;
%$\Bx(1), \Bx(2), ... \Bx(I)$, 
%each with dimensions $L \times 1$
%(the value of $I$ is determined by network parameters). 
%Transmission vectors 
%which
%are then transmitted in a TDMA fashion
%, using the array of $L$ antennas
where $I$ is a parameter determined by the delivery algorithm. After $\Bx(i)$ is transmitted, user $k$ receives 
$ y_k(i) = \Bh_k^T \Bx(i) + z_k(i)$,
%\begin{equation}
%    y_k(i) = \Bh_k^T \Bx(i) + z_k(i) \; ,
%\end{equation}
where $\Bh_k \in \mathbb{C}^L$ is the channel vector
% from $L$ transmitting antennas to 
for user $k$ and $z_k(i) \sim \mathbb{C}\CN(0,N_0)$ represents the observed noise at user $k$ during transmission interval $i$.
%
%The placement and delivery algorithms, jointly referred to as the caching scheme, should be designed such that each user $k \in [K]$ can decode its requested file $W(k)$, using $\CZ(k)$ (its locally cached data), together with $y_k(i), i \in [I]$ (data it has received from the channel). 
%Delivery time $T$ is defined as the time required for all users to decode their requested files.
%Denoting the worst case delivery time (with respect to $\CD$) as $T^*$, the goal is to find caching schemes resulting in smaller $T^*$. Following the common practice in the literature, we assume each user has requested a different file, in order to calculate $T^*$.
%Moreover, we also use $A \equiv W(1)$, $B \equiv W(2)$, etc., for notation simplicity.

%As cache placement is done without any knowledge of file request probabilities, an efficient strategy is to store equal-sized data portions of all files in the cache memory of each user. Thereby, every user has $\frac{M}{N}$ of each file in its cache memory, and
As $\nicefrac{M}{N}$ of every file is available at $\CZ(k)$, user $k$ needs to get $(1 - \nicefrac{M}{N})$ of
%should receive 
%the rest $(1 - \frac{M}{N})$ of 
$W(k)$
%its requested 
%file 
from the server.
%This results in a total data size of $K(1-\frac{M}{N})$ to be transmitted over the channel. 
Let us define the
%global cache ratio (
coded caching gain as
%) $t$ as the total cache size in the network normalized by the number of files, i.e. 
$t = \nicefrac{KM}{N}$
%; 
and assume $t$ is an integer.
Defining delivery time $T$ as the time required for all users to decode their requested files,
%Then the 
the
effective communication rate
%, denoted by $R$, 
is % defined as
$R = \nicefrac{K(1-\nicefrac{t}{K})}{T}$.
%\begin{equation}
%\label{eq:sum_rate_def}
%    R = \frac{K(1-\frac{t}{K})}{T} \; .
%\end{equation}
%
For a simple channel with capacity one
%(normalized) 
data unit per channel use, $R$ represents how many users simultaneously benefit from each transmission. 
%Denoting the effective rate under this simple channel condition as $R_s$, we 
We
use the term Degree of Freedom (DoF) equivalent to %$R_s$. 
$R$ is such a condition.
%DoF is one of the parameters used in the literature for comparing various caching schemes.

%\subsection{Coded Caching (CC)}
If $L=1$, a trivial scheme (unicasting every missing data part) achieves DoF of one.
%, by unicasting every missing data part. 
Interestingly, CC enables DoF of $t+1$ to be achieved, for the same setup \cite{maddah2014fundamental}.
%; through multicasting of carefully designed codewords during each transmission interval. Technically, 
During placement phase, each file $W$ is split into $\binom{K}{t}$ smaller parts $W_{\CT}$, where $\CT \subseteq [K]$, $|\CT| = t$; and 
$\CZ(k) = \{ W_{\CT} \mid W \in \CF, \CT \ni k \}$.
%\begin{equation}
%    \CZ(k) = \{ W_{\CT} \mid W \in \CF, \CT \ni k \} \; .
%\end{equation}
%
%During the delivery phase, 
For delivery, for every $\CS \subseteq [K], |\CS| = t+1$,
%a codeword 
$X(\CS)$ is built as
\begin{equation}
    X(\CS) = \bigoplus\limits_{k \in \CS} W_{\CS \backslash \{k\}}(k) \; ,
\end{equation}
where $\oplus$ denotes the bit-wise XOR operation; and each $X(\CS)$ is broadcast in a separate time interval. %Every user $k \in \CS$ can then use $\CZ(k)$ to remove unwanted terms; and decode $W_{\CS \backslash \{k\}} (k)$ interference-free.
Following the same structure, in \cite{shariatpanahi2016multi} it is shown that in a multi-server scenario with $L$ servers, the DoF of $t+L$ is achievable. This result is then extended in \cite{shariatpanahi2018physical} to an $L$-antenna MISO setup.

\subsection{The Subpacketization Effect}
%Codeword creation procedure of 
Subpacketization $P$ is defined as the number of smaller parts each file should be split into.
%Any CC scheme requires all the files to be split into $P$ smaller parts, where $P$ is known as subpacketization. 
The original CC scheme \cite{maddah2014fundamental} requires $P = \binom{K}{t}$; which means $P$ grows exponentially with $K$, if $t$ also scales with $K$ (polynomially, if $t$ is fixed). For the multi-server scheme \cite{shariatpanahi2016multi}, the growth in $P$ is even worse by a multiplicative factor of $\binom{K-t-1}{L-1}$. This makes the CC implementation infeasible, even for moderate values of $K$ \cite{lampiris2018adding}.
%; and hence reducing subpacketization (without any loss in DoF) has been thoroughly studied in the literature, for both single- and multi-antenna communication setups.

%For single-antenna setups, reducing $P$ is systematically studied in \cite{yan2017placement}, where Placement Delivery Arrays (PDA) are used to show that the original scheme of \cite{maddah2014fundamental} is in fact optimal, among a class of symmetric schemes known as $g$-regular PDA. The same structure is then used in \cite{yan2017placementb}, to show that for $t$ scaling with $K$, a PDA resulting in linear $P$ does not exist. 
%Finally, in \cite{shangguan2018centralized} hypergraphs are used to design a CC scheme with sub-exponentially growing subpacketization.

Interestingly, multi-antenna setups enable huge reductions in $P$ to be achieved, without any loss in DoF.
%Multi-antenna CC schemes are more flexible for reducing $P$. 
In \cite{lampiris2018adding} %it is shown that any single-antenna CC scheme can be \textit{elevated} for multi-antenna setups. 
a structure for \textit{elevating} single-antenna CC schemes for multi-antenna setups is introduced. The resulting CC scheme would then require $P' = g(\frac{K}{L},\frac{t}{L})$; if for the original one $P = g(K,t)$ for some function $g$. However, this structure incurs DoF loss (up to a factor of 2), if either $\nicefrac{K}{L}$ or $\nicefrac{t}{L}$ is non-integer. In \cite{salehi2019coded} a CC scheme with $P=K(t+L)$ is introduced, for networks with $L \ge t$.
%, applicable to any set of network parameters as long as $L \ge t$, results in linear $P$ with respect to $K$, if $t$ is not scaling with $K$.
In \cite{salehi2019subpacketization} it is shown that if $K = t+L$, $P$ can be selected freely among a set of predefined values.
%; enabling flexible subpacketization for a specific set of network parameters.

\subsection{The Beamformer Effect}
%Considering real-world wireless networks, DoF is not the best performance metric; as it is defined for the case of channel capacity being one (normalized) data unit per channel use, which is not valid due to fading and free movement of users. To address this issue, symmetric rate is considered as the metric in \cite{shariatpanahi2017multi,shariatpanahi2018physical}; and ZF beamformers are used to create the transmission vectors.
%
Zero-forcing beamformers (ZF), used in the original multi-antenna CC scheme \cite{shariatpanahi2018physical},
%The original multi-antenna scheme \cite{shariatpanahi2018physical} assumes ZF beamformers.
are shown in \cite{tolli2018multicast} to result in poor rate in low-SNR communications.
%; and one needs to use optimized beamformers to achieve higher data rates.
%that ZF beamformers result in poor achievable rate at finite-SNR communications; and one needs optimized beamformers to improve the performance at this regime.
In \cite{tolli2017multi} interesting methods % based on two design parameters $\alpha$ and $\beta$ are introduced, to reduce 
for reducing optimized beamformer design complexity are proposed. They incur either DoF loss or increased $P$, however.

\subsection{Our Model}
Following \cite{salehi2019subpacketization}, we assume $K = t+L$; and
%In \cite{salehi2019subpacketization} it is shown that using optimized beamformers, a trade-off does exist between $P$ and the symmetric rate, such that larger $P$ results in increased rate. We study the same structure under different beamformer design policies and show that counter-intuitively, if the beamformer structure is simpler, increasing $P$ in fact reduces the achievable rate, at finite-SNR. Using mathematical and numerical analysis, we investigate the cause of such behavior; and explore the interaction between subpacketization selection and beamformer design policy.
build $\CZ(k)$ using a placement matrix $\BV$, which is 
%use placement matrices to build $\CZ(k)$. A placement matrix $\BV$ is 
a $P \times K$ binary matrix with $\sum_p \BV[p,k] = t, \forall k \in [K]$ and $\sum_k \BV[p,k] = \nicefrac{Pt}{K}, \forall p \in [P]$. Each file $W$ is split into $P$ smaller parts $W_p$; and% the cache placement is done as
\begin{equation}
    \CZ(k) = \{ W_p \mid \BV[p,k] = 1; \; \forall p \in [P], \forall W \in \CF \} \; .
\end{equation}
For delivery, for each $\CS \subseteq [K], |\CS| = t+1$ we define% the corresponding packet set of $\CS$ as
\begin{equation}
    \Phi(\CS) = \{ p \in [P] \mid \BV[p,k] = 0 , \; \forall k \in [K] \backslash \CS \} \; ,
\end{equation}
and build codeword $X(\CS)$ as
\begin{equation}
    X(\CS) = \bigoplus \limits_{\substack{k \in \CS \\ p \in \Phi(\CS)}} (1-\BV[p,k])W_p(k) \; .
\end{equation}
As $K=t+L$, one can send all codewords $X(\CS)$ %can be sent with a 
in a single interval. The transmission vector is built as $\Bx = \sum_{\CS} \Bw_{\CS} X(\CS)$; 
%\begin{equation}
%    \Bx = \sum_{\CS} \Bw_{\CS} X(\CS) \; ,
%\end{equation}
where $\Bw_{\CS} \in \mathbb{C}^L$ is the beamforming vector associated to $X(\CS)$. %, built according to the beamformer structure. 
Let us denote the total power constraint as $Po$; and the number of $\CS$ sets for which $\Phi(\CS) \neq \varnothing$ as $n(\BV)$. We study three different beamformer structures:
\begin{itemize}[leftmargin=*]
    \item[-] \textbf{EP:} ZF with uniform power allocation, for which
    \begin{equation}
    \label{eq:beamformer_EP}
        \Bw_{\CS} = \sqrt{\frac{Po}{n(\BV)}} \times \Bu_{\CS} \; ,
    \end{equation}
    where $\Bu_{\CS}$ is the ZF vector associated with $\CS$, built such that $\| \Bu_{\CS} \| = 1$ and $\Bh_k^T \Bu_{\CS} = 0$, $\forall k \in [K] \backslash \CS$.
    \item[-] \textbf{PL:} ZF with optimal power allocation, for which
    \begin{equation}
    \label{eq:beamformer_PL}
        \Bw_{\CS} = \sqrt{\alpha_{\CS} Po} \times \Bu_{\CS} \; ,
    \end{equation}
    where %$\Bu_{\CS}$ is the ZF vector and 
    $\alpha_{\CS}$ is the power coefficient of $X(\CS)$, selected such that $R$ is maximized and $\sum_{\CS} \alpha_{\CS} = 1$.
    \item[-] \textbf{BF:} optimized beamformer; for which
    \begin{equation}
    \label{eq:beamformer_BF}
        \Bw_{\CS} = \sqrt{\alpha_{\CS} Po} \times \Bv_{\CS} \; ,
    \end{equation}
    where %$\alpha_{\CS}$ and 
    $\Bv_{\CS}$ is the beamformer vector of $X(\CS)$, designed such that $R$ is maximized and $\| \Bv_{\CS} \| = 1$.
\end{itemize}

According to \cite{salehi2019subpacketization}, after $\Bx$ is transmitted, %the transmission of $\Bx$ is concluded,
%; every user $k \in [K]$ is able to decode all its missing data parts. In fact, 
all unwanted terms at user $k$ are either zero-forced (suppressed, in case of BF), or removed by $\CZ(k)$. Consequently, user $k$ has to decode its requested data parts
%the missing data parts (useful terms)
from a multiple access channel (MAC) of size $m(\BV) = P - \frac{Pt}{K} = \frac{PL}{K}$.
%as the number of useful terms in the MAC channel (MAC size) for the placement matrix $\BV$, we have
%\begin{equation}
%    m(\BV) = P - \frac{Pt}{K} = \frac{P(K-t)}{K} = \frac{PL}{K} \; .
%\end{equation}
%
%From communication theory, we know that 
%The rate region of a MAC channel with size $j$ has $2^j-1$ constraints. However, 
Let us use $r_k^j$ to denote the rate associated with the $j$-th term, $j \in [m(\BV)]$, at the MAC channel of user $k$. All terms in the MAC channel should be decoded, for user $k$ to receive $W(k)$ successfully. This means $r_k = \min_j r_k^j$, where $r_k$ is the perceived rate at user $k$. Let us define the symmetric rate as $r_s = \min r_k, k \in [K]$; and use $\mathrm{SINR}_k^j$ to denote the SINR value of $r_k^j$, at user $k$. Then the rate optimization problem is (note that the power constraint is implicitly present in the beamformer structure):
%All missing data parts should be available at user $k$, for $W(k)$ to be delivered successfully.
%So, using $r_k^j$ to denote the rate associated with the $j$-th term, $j \in [m(\BV)]$, at the MAC channel of user $k$
%Hence the perceived rate at user $k$, denoted by $r_k$, is limited to the minimum rate of all terms in its MAC channel. Defining the symmetric rate as $r_s = \min r_k, k \in [K]$, the rate optimization problem can be written as (for notation simplicity, we ignore explicit representation of power constraint)
%
\begin{equation}
\label{eq:optimization_problem_main}
\begin{aligned}
    \max \quad &r_s = \min\limits_{k \in [K]} \min\limits_{j \in [m(\BV)]} r_k^j \\
    s.t. \quad %&r_k = \min\limits_{j \in [m(\BV)]} r_k^j \qquad \qquad \qquad \quad \forall k \in [K] \; , \\
    &\sum_{j \in \CJ} r_k^j \le \log (1 + \sum_{j \in \CJ} \mathrm{SINR}_k^j) \\
    &\qquad \qquad \forall k \in [K]; \; \forall \CJ \subseteq[m(\Bv)], \CJ \neq \varnothing \; .
\end{aligned}
\end{equation}
%where $r_k^j$ is the rate associated with the $j$-th term, $j \in [m(\BV)]$, at the MAC channel of user $k$; and 
%where $\mathrm{SINR}_k^j$ is the signal to noise plus interference value associated with $r_k^j$, at user $k$; and power constraint is implicitly present in the beamformer structure.

%The symmetric rate $r_s$ is the rate with which each user can decode all its requested data parts simultaneously. 
In order to compare various schemes, we use the total delivery time $T$. As the size of each data part is $\nicefrac{1}{P}$ and all parts are decoded simultaneously, we have $T = \frac{1}{P} \frac{1}{r_s}$.
%\begin{equation}
%    \label{eq:delivery_time_main}
%    T = \frac{1}{P} \frac{1}{r_s} \; .
%\end{equation}

\subsection{Motivation}
\label{sec:motivation}
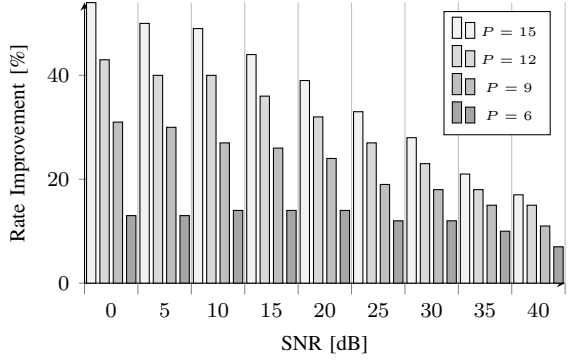
\begin{figure}[t]
    \centering
    %\resizebox{0.9\columnwidth}{!}
    {%
   
    \begin{tikzpicture}
    \begin{axis}
    [
    width = 0.9\columnwidth,
    height = 0.6\columnwidth,
    % put axis lines at left and bottom
    axis lines = left,
    % control axis labels
    xlabel = \smaller {SNR [dB]},
    ylabel = \smaller {Rate Improvement [\%]},
    ylabel near ticks,
    % control legend position
    legend pos = north east,
    legend style = {cells={align=center}},
    % control size of tick marks (10,20,30,etc)
    ticklabel style={font=\smaller},
    % control major grids
    %grid=both,
    %major grid style={line width=.2pt,draw=gray!30},
    % control minor grids
    %grid style={line width=.1pt, draw=gray!10},
    %minor tick num=5,
    ybar interval=0.7,
    ymin = 0,
    ]
    
    \addplot
    [black,fill = gray!10]
    table
    [x=SNR,y=P15]
    {Figures/Data/BF_Bars.txt};
    \addlegendentry{\tiny $P=15$}

    \addplot 
    [black,fill = gray!30]
    table
    [x=SNR,y=P12]
    {Figures/Data/BF_Bars.txt};
    \addlegendentry{\tiny $P=12$}
    
    \addplot 
    [black,fill = gray!50]
    table
    [x=SNR,y=P9]
    {Figures/Data/BF_Bars.txt};
    \addlegendentry{\tiny $P=9$}
    
    \addplot 
    [black,fill = gray!70]
    table
    [x=SNR,y=P6]
    {Figures/Data/BF_Bars.txt};
    \addlegendentry{\tiny $P=6$}
    
    %\addplot 
    %[
    %black,
    %fill = gray!90,
    %]
    %table
    %[x=SNR,y=P3ZF]
    %{Data/K6t2_Bar.txt};
    %\addlegendentry{\smaller $P=3$, ZF}
    
    \end{axis}
    \end{tikzpicture}
    }
    \caption{Rate Improvement over $P=3$ - BF beamformer}
    \label{fig:bf_bar}
\end{figure}

\begin{figure}[t]
    \centering
    %\resizebox{0.9\columnwidth}{!}
    {%
   
    \begin{tikzpicture}
    \begin{axis}
    [
    width = 0.9\columnwidth,
    height = 0.6\columnwidth,
    % put axis lines at left and bottom
    axis lines = left,
    % control axis labels
    xlabel = \smaller {SNR [dB]},
    ylabel = \smaller {Rate Improvement [\%]},
    ylabel near ticks,
    % control legend position
    legend pos = south east,
    legend style = {cells={align=center}},
    % control size of tick marks (10,20,30,etc)
    ticklabel style={font=\smaller},
    % control major grids
    %grid=both,
    %major grid style={line width=.2pt,draw=gray!30},
    % control minor grids
    %grid style={line width=.1pt, draw=gray!10},
    %minor tick num=5,
    ybar interval=0.7,
    ymin = -30,
    ymax = 32
    ]
    
    \addplot
    [black,fill = gray!10]
    table
    [x=SNR,y=P15]
    {Figures/Data/PL_Bars.txt};
    \addlegendentry{\tiny $P=15$}

    \addplot 
    [black,fill = gray!30]
    table
    [x=SNR,y=P12]
    {Figures/Data/PL_Bars.txt};
    \addlegendentry{\tiny $P=12$}
    
    \addplot 
    [black,fill = gray!50]
    table
    [x=SNR,y=P9]
    {Figures/Data/PL_Bars.txt};
    \addlegendentry{\tiny $P=9$}
    
    \addplot 
    [black,fill = gray!70]
    table
    [x=SNR,y=P6]
    {Figures/Data/PL_Bars.txt};
    \addlegendentry{\tiny $P=6$}
    
    %\addplot 
    %[
    %black,
    %fill = gray!90,
    %]
    %table
    %[x=SNR,y=P3ZF]
    %{Data/K6t2_Bar.txt};
    %\addlegendentry{\smaller $P=3$, ZF}
    
    \end{axis}
    \end{tikzpicture}
    }
    \caption{Rate Improvement over $P=3$ - PL beamformer}
    \label{fig:pl_bar}
\end{figure}
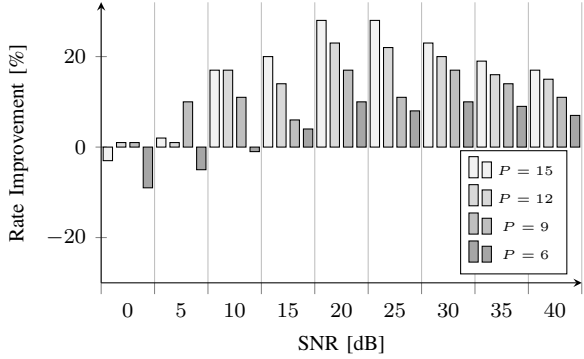

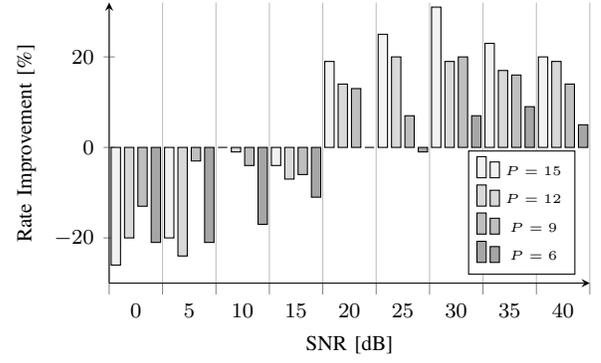
\begin{figure}[t]
    \centering
    %\resizebox{0.9\columnwidth}{!}
    {%
   
    \begin{tikzpicture}
    \begin{axis}
    [
    width = 0.9\columnwidth,
    height = 0.6\columnwidth,
    % put axis lines at left and bottom
    axis lines = left,
    % control axis labels
    xlabel = \smaller {SNR [dB]},
    ylabel = \smaller {Rate Improvement [\%]},
    ylabel near ticks,
    % control legend position
    legend pos = south east,
    legend style = {cells={align=center}},
    % control size of tick marks (10,20,30,etc)
    ticklabel style={font=\smaller},
    % control major grids
    %grid=both,
    %major grid style={line width=.2pt,draw=gray!30},
    % control minor grids
    %grid style={line width=.1pt, draw=gray!10},
    %minor tick num=5,
    ybar interval=0.7,
    ymin = -30,
    ymax = 32
    ]
    
    \addplot
    [black,fill = gray!10]
    table
    [x=SNR,y=P15]
    {Figures/Data/EP_Bars.txt};
    \addlegendentry{\tiny $P=15$}

    \addplot 
    [black,fill = gray!30]
    table
    [x=SNR,y=P12]
    {Figures/Data/EP_Bars.txt};
    \addlegendentry{\tiny $P=12$}
    
    \addplot 
    [black,fill = gray!50]
    table
    [x=SNR,y=P9]
    {Figures/Data/EP_Bars.txt};
    \addlegendentry{\tiny $P=9$}
    
    \addplot 
    [black,fill = gray!70]
    table
    [x=SNR,y=P6]
    {Figures/Data/EP_Bars.txt};
    \addlegendentry{\tiny $P=6$}
    
    %\addplot 
    %[
    %black,
    %fill = gray!90,
    %]
    %table
    %[x=SNR,y=P3ZF]
    %{Data/K6t2_Bar.txt};
    %\addlegendentry{\smaller $P=3$, ZF}
    
    \end{axis}
    \end{tikzpicture}
    }
    \caption{Rate Improvement over $P=3$ - EP beamformer}
    \label{fig:ep_bar}
\end{figure}

%The effect of $P$ value on the effective rate $R$ with BF structure is studied in \cite{salehi2019subpacketization}. 
The results 
of \cite{salehi2019subpacketization} 
indicate 
that using BF structure, 
$R$ is improved as $P$ is increased; and this effect is more prominent at lower SNR. The improvement in $R$ for various $P$ values with respect to $P=3$ is depicted in Figure \ref{fig:bf_bar}, for a network with $K=6, t=2, L=4$. Consequently, for the same setup, we have plotted the results for PL and EP structures, in Figures \ref{fig:pl_bar} and \ref{fig:ep_bar} respectively. Clearly, at low-SNR regime, the performance of $P>3$ is worse than $P=3$ for both structures. Specially for EP, at $0 \mathrm{dB}$ the performance is degraded as $P$ is increased from 6 to 15. This indicates a difference in how $P$ affects the performance, for various beamformer structures.
%Counter-intuitively, changing the beamformer structure has reversed the results of \cite{salehi2019subpacketization}; i.e. increasing $P$ has degraded $R$, with degradation being more severe for the EP policy.

\section{Mathematical Analysis}
\label{sec:analysis}
%To investigate the effect of beamformer structure, 
%We first consider a simple network with $K=4$, $t=2$, $L=2$; and then provide results for more general setups. 
% and provide closed-form performance expressions for this network. 
Using the low-SNR approximation $\log (1 + \mathrm{SINR}) \simeq \mathrm{SINR}$
%\begin{equation}
%\label{eq:finiteSNR_approx}
%    \log (1 + \mathrm{SINR}) \simeq \mathrm{SINR} \; ,
%\end{equation}
(using Taylor expansion of $\log(1+x)$, as $x \rightarrow 0$; $\log$ is the natural logarithm), %eliminates all sum-rate constraints of 
the optimization problem \eqref{eq:optimization_problem_main} is reduced to
\begin{equation}
\label{eq:optimization_problem_simple}
\begin{aligned}
    \max \quad &r_s = \min\limits_{k \in [K]} \min\limits_{j \in [m(\BV)]} r_k^j \\
    s.t. \quad %&r_k =  \qquad \qquad \qquad \quad \forall k \in [K] \; , \\
    &r_k^j \le \mathrm{SINR}_k^j \qquad \; \; \forall k \in [K]; \; \forall j \in [m(\BV)] \; .
\end{aligned}
\end{equation}
Note that as $P$ becomes larger, number of constraints being removed as a result of approximation grows exponentially; and hence the approximation effect on $r_s$ is more prominent.
%
%Note that the effect of the finite-SNR approximation %\eqref{eq:finiteSNR_approx} 
%on $r_s$ is more prominent as $P$ becomes larger. This is because for larger $P$, the number of constraints being removed as a result of approximation grows exponentially.

We first consider a simple network with $K=4$, $t=2$, $L=2$; and then study more general setups. For the simple network,
we assume $\CD = \{A,B,C,D \}$; and calculate $T$ for $P \in \{2,4,6 \}$ and beamformer structures EP, PL, BF.
%$\gamma \in \{\mathrm{EP},\mathrm{PL},\mathrm{BF} \}$. 
We use $\Bu_k = \Bu_{[4]\backslash\{k\}}$, $\Bw_k = \Bw_{[4]\backslash\{k\}}$ and $\alpha_k = \alpha_{[4]\backslash\{k\}}$, $k \in [4]$, for notation simplicity. Placement matrices for $P=2,4$ are $\BV_1$, $\BV_2$ as mentioned in \eqref{eq:placement_matrix_simple_net}; and for $P=6$, $\BV_3$ is column-wise concatenation of $\BV_1$ and $\BV_2$.
\begin{equation}
\label{eq:placement_matrix_simple_net}
\BV_1 =
    \begin{bmatrix}
    1 & 0 & 1 & 0 \\
    0 & 1 & 0 & 1
    \end{bmatrix} \, ,
\BV_2 =
    \begin{bmatrix}
    1 & 1 & 0 & 0 \\
    0 & 1 & 1 & 0 \\
    0 & 0 & 1 & 1 \\
    1 & 0 & 0 & 1 \\
    \end{bmatrix} \, .
\end{equation}
%For $P=6$, $\BV_3$ is column-wise concatenation of $\BV_1$ and $\BV_2$. 

\subsection{EP Beamformer, $P=2$}
\label{section:case_ep_p2}
Based on $\BV_1$, we have 
$\CZ(1) = \CZ(3) = \{ W_1 \mid W \in \CF \}$, $\CZ(2) = \allowbreak \CZ(4) = \{ W_2 \mid W \in \CF \}$. 
Using EP beamformer \eqref{eq:beamformer_EP}, 
the transmission vector is built as
\begin{equation}
\label{eq:trans_vector_EP_P2}
\begin{aligned}
    \Bx = \sqrt{\nicefrac{Po}{4}} \big[ A_2 \Bu_{3} +  B_1 \Bu_{4}
    +  C_2 \Bu_{1} + D_1 \Bu_{2} \big] \; .
\end{aligned}
\end{equation}
%According to the EP beamformer definition, 
Based on ZF definition, the third term %in \eqref{eq:trans_vector_EP_P2} 
is nulled at user 1; i.e. %and so we have
\begin{equation}
\label{eq:rec_signal_EP_P2}
\begin{aligned}
    y_1 = \sqrt{\nicefrac{Po}{4}} \big[ A_2 \Bh_1^T \Bu_{3} +  B_1 \Bh_1^T \Bu_{4}
    +  D_1 \Bh_1^T \Bu_{2} \big] + z_1 \; .
\end{aligned}
\end{equation}
Now user 1 can reconstruct and remove the second and third terms in \eqref{eq:rec_signal_EP_P2} using %its cache contents 
$\CZ(1)$; and decode %its requested data part 
$A_2$ with SINR
\begin{equation}
    \mathrm{SINR}_1^1 = \frac{Po}{4} \frac{|\Bh_1^T \Bu_{\{1,2,4\}}|^2}{N_0} \; .
\end{equation}

Following the same procedure for all users, the symmetric rate can be calculated as
\begin{equation}
    r_s^{EP,\BV_1} = \frac{Po}{4N_0} \min \CL_{EP}^{\BV_1} \; ,
\end{equation}
where $\CL_{EP}^{\BV_1} = \big\{ |\Bh_1^T \Bu_{3}|^2, |\Bh_2^T \Bu_{4}|^2, |\Bh_3^T \Bu_{1}|^2, |\Bh_4^T \Bu_{2}|^2 \big\}$.
%\begin{equation}
%    \CL_{EP}^{\BV_1} = \big\{ |\Bh_1^T \Bu_{3}|^2, |\Bh_2^T \Bu_{4}|^2, |\Bh_3^T \Bu_{1}|^2, |\Bh_4^T \Bu_{2}|^2 \big\} \; .
%\end{equation}
%
For the total delivery time we have
\begin{equation}
    T_{EP}^{\BV_1} = \frac{1}{2} \frac{1}{r_s^{EP,\BV_1}} = \frac{1}{2} \frac{N_0}{Po} \frac{4}{\min \CL_{EP}^{\BV_1}} \; .
\end{equation}

\subsection{EP Beamformer, $P=4$}
Following the same procedure in $P=2$, $\Bx$ is built as
\begin{equation}
\begin{aligned}
    \Bx = \sqrt{\nicefrac{Po}{4}} \big[ &(A_2 \oplus C_1) \Bu_{4} + (A_3 \oplus C_4) \Bu_{2} \\ + &(B_3 \oplus D_2) \Bu_{1} + (B_4 \oplus D_1) \Bu_{3} \big] \; ,
\end{aligned}
\end{equation}
and user 1 receives
\begin{equation}
\begin{aligned}
        y_1 = \sqrt{\nicefrac{Po}{4}} \big[ (&A_2 \oplus C_1) \Bh_1^T \Bu_{4} 
        + (A_3 \oplus C_4) \Bh_1^T \Bu_{2} \\
        &+ (B_4 \oplus D_1) \Bh_1^T \Bu_{3} \big] + z_1 \; .
\end{aligned}
\end{equation}
User 1 can reconstruct and remove the third term using $\CZ(1)$, and decode $(A_2 \oplus C_1)$ and $(A_3 \oplus C_4)$ 
%from a MAC channel of size 2, 
with SINR values
\begin{equation}
    \mathrm{SINR}_1^1 = \frac{Po}{4} \frac{|\Bh_1^T \Bu_{3}|^2}{N_0} , \; \mathrm{SINR}_1^2 = \frac{Po}{4} \frac{|\Bh_1^T \Bu_{2}|^2}{N_0} \; .
\end{equation}
Next, user 1 again uses $\CZ(1)$ to extract its requested terms $A_2, A_3$. Following the same procedure for all users, we have
\begin{equation}
    r_s^{EP,\BV_2} = \frac{Po}{4N_0} \min \CL_{EP}^{\BV_2} \; ,
\end{equation}
where
\begin{equation}
    \begin{aligned}
        \CL_{EP}^{\BV_2} = \big\{ &|\Bh_1^T \Bu_{4}|^2, |\Bh_1^T \Bu_{2}|^2, |\Bh_2^T \Bu_{1}|^2, |\Bh_2^T \Bu_{3}|^2, \\
        &|\Bh_3^T \Bu_{4}|^2, |\Bh_3^T \Bu_{2}|^2, |\Bh_4^T \Bu_{3}|^2, |\Bh_4^T \Bu_{1}|^2 \big\} \; .
    \end{aligned}
\end{equation}
Then for the total delivery time we have
\begin{equation}
    T_{EP}^{\BV_2} = \frac{1}{4} \frac{1}{r_s^{EP,\BV_2}} = \frac{1}{4} \frac{N_0}{Po} \frac{4}{\min \CL_{EP}^{\Bv_2}} \; .
\end{equation}

\subsection{EP Beamformer, $P=6$}
Following the same procedure, 
%as $P=2,4$, $\Bx$ is built as
%\begin{equation}
%\begin{aligned}
%    \Bx = \sqrt{\nicefrac{Po}{4}} &\big[ (A_2 \oplus C_1 \oplus B_5) \Bu_{4} + (A_3 \oplus C_4 \oplus D_5) \Bu_{2} \\
%     + (&B_3 \oplus D_2 \oplus C_6) \Bu_{1} + (B_4 \oplus D_1 \oplus A_6) \Bu_{3} \big] \; .
%\end{aligned}
%\end{equation}
%Now 
each user has to decode its requested terms through a MAC channel of size 3; and %the symmetric rate is
\begin{equation}
    r_s^{EP,\BV_3} = \frac{Po}{4N_0} \min \CL_{EP}^{\BV_3} \; ,
\end{equation}
where $\CL_{EP}^{\BV_3}$ is defined as
\begin{equation}
    \begin{aligned}
        \CL_{EP}^{\BV_3} = \big\{ &|\Bh_1^T \Bu_{4}|^2, |\Bh_1^T \Bu_{2}|^2, |\Bh_1^T \Bu_{3}|^2, |\Bh_2^T \Bu_{1}|^2, \\
        &|\Bh_2^T \Bu_{3}|^2, |\Bh_2^T \Bu_{4}|^2, |\Bh_3^T \Bu_{4}|^2, |\Bh_3^T \Bu_{2}|^2, \\
        &|\Bh_3^T \Bu_{1}|^2, |\Bh_4^T \Bu_{3}|^2, |\Bh_4^T \Bu_{1}|^2, |\Bh_4^T \Bu_{2}|^2 \big\} \; .
    \end{aligned}
\end{equation}
The total delivery time can then be calculated as
\begin{equation}
    T_{EP}^{\BV_3} = \frac{1}{6} \frac{1}{r_s^{EP,\BV_3}} = \frac{1}{6} \frac{N_0}{Po} \frac{4}{\min \CL_{EP}^{\Bv_3}} \; .
\end{equation}

\subsection{PL Beamformer}
According to \eqref{eq:beamformer_PL}, using PL structure instead of EP, the power coefficient $\nicefrac{Po}{4}$ is replaced with $\alpha_{\CS}Po$. This causes $r_s$ to be a function of $\alpha_{\CS}$; which should be optimized in order to maximize $r_s$. Starting from the case $P=2$, we define
\begin{equation*}
    \begin{aligned}
        \CL_{PL}^{\BV_1} = \big\{ \alpha_{3} |\Bh_1^T \Bu_{3}|^2, \alpha_{4} |\Bh_2^T \Bu_{4}|^2, \alpha_{1} |\Bh_3^T \Bu_{1}|^2, \alpha_{2} |\Bh_4^T \Bu_{2}|^2 \big\} \; .
    \end{aligned}
\end{equation*}
The symmetric rate is then calculated as
\begin{equation}
    \begin{aligned}
        r_s^{PL,\BV_1} &= \frac{Po}{N_0} \max\limits_{\alpha_{\CS}} \min \CL_{PL}^{\BV_1} \; , \\
    \end{aligned}
\end{equation}
and the delivery time will be
\begin{equation}
    \begin{aligned}
        T_{PL}^{\BV_1} &= \frac{1}{2} \frac{1}{r_s^{PL,\BV_1}} = \frac{1}{2} \frac{N_0}{Po} \frac{1}{\max\limits_{\alpha_{\CS}} \min \CL_{PL}^{\BV_1}} \; . \\
    \end{aligned}
\end{equation}
Delivery times for $P=4,6$ are calculated similarly.

\subsection{BF Beamformer}
Using BF instead of EP, $\nicefrac{Po}{4}$ is replaced with $\alpha_{\CS}Po$ and $\Bv_{\CS}$ is used instead of $\Bu_{\CS}$. This causes $r_s$ to be a function of both $\alpha_{\CS}$ and $\Bv_{\CS}$; and SINR terms to include interference from unwanted terms. Considering the case $P=2$, we have
\begin{equation*}
\label{eq:trans_vector_BF_P2}
\begin{aligned}
    \Bx \! = \! \sqrt{Po} \big[ A_2 \sqrt{\alpha_{3}} \Bv_{3} +  B_1 \sqrt{\alpha_{4}} \Bv_{4} + C_2 \sqrt{\alpha_{1}} \Bv_{1} + D_1 \sqrt{\alpha_{2}} \Bv_{2} \big],
\end{aligned}
\end{equation*}
and user 1 receives
\begin{equation}
\label{eq:rec_signal_BF_P2}
\begin{aligned}
    y_1 = \sqrt{Po} &\big[ A_2 \sqrt{\alpha_{3}} \Bh_1^T \Bv_{3} +  B_1 \sqrt{\alpha_{4}} \Bh_1^T \Bv_{4} \\
    +& C_2 \sqrt{\alpha_{1}} \Bh_1^T \Bv_{1} + D_1  \sqrt{\alpha_{2}} \Bh_1^T \Bv_{2} \big] + z_1 \; ,
\end{aligned}
\end{equation}
from which it can reconstruct and remove the second and fourth terms, using $\CZ(1)$.
The third term appears as interference however; and hence for decoding $A_2$ at user 1 we have
\begin{equation}
    \mathrm{SINR}_1^1 = \alpha_{3} Po \frac{|\Bh_1^T \Bv_{3}|^2}{|\Bh_1^T \Bv_{1}|^2 + N_0} \; .
\end{equation}
the symmetric rate can then be calculated as
\begin{equation}
    r_s^{BF,\BV_1} = \frac{Po}{N_0} \max\limits_{\alpha_{\CS}, \Bv_{\CS}} \min \CL_{BF}^{\BV_1} \; ,
\end{equation}
where $\CL_{BF}^{\BV_1}$ is defined as
\begin{equation}
\begin{aligned}
        \CL_{BF}^{\BV_1} = \big\{ &\frac{\alpha_{3} N_0 |\Bh_1^T \Bv_{3}|^2}{\alpha_{1} |\Bh_1^T \Bv_{1}|^2 + N_0}, \frac{\alpha_{4} N_0 |\Bh_2^T \Bv_{4}|^2}{\alpha_{2} |\Bh_2^T \Bv_{2}|^2 + N_0}, \\
        &\frac{\alpha_{1} N_0 |\Bh_3^T \Bv_{1}|^2}{\alpha_{3} |\Bh_3^T \Bv_{3}|^2 + N_0}, \frac{\alpha_{2} N_0 |\Bh_1^T \Bv_{2}|^2}{\alpha_{4} |\Bh_4^T \Bv_{4}|^2 + N_0} \big\} \; .
\end{aligned}
\end{equation}
Finally, for delivery time we have
\begin{equation}
    T_{BF}^{\BV_1} = \frac{1}{2} \frac{1}{r_s^{BF,\Bv_1}} = \frac{1}{2} \frac{N_0}{P} \frac{1}{\max\limits_{\alpha_{\CS}, \Bv_{\CS}} \min \CL_{BF}^{\BV_1}} \; .
\end{equation}
Delivery times for $P=4,6$ are calculated similarly. Defining
\begin{equation}
    \begin{gathered}
        \Tilde{\CS} = \{ \CS \subseteq [K] \, ; \, |\CS| = t+1 \} \; , \\
        \CS(k) = \big\{ \CS \in \Tilde{\CS} \; \big| \;  k \in \CS ; \; \exists p \in \Phi(\CS): \BV[p,k] = 0 \big\} \; , \\
        \overline{\CS}(k) = \big\{ \CS \in \Tilde{\CS} \; \big| \; k \not\in \CS \big\} \; ,
    \end{gathered}
\end{equation}
%$\Tilde{\CS} = \{ \CS \subseteq [K] \, ; \, |\CS| = t+1 \}$, $\CS(k) = \{ \CS \in \Tilde{\CS} \; \big| \;  k \in \CS ; \; \exists p \in \Phi(\CS): \BV[p,k] = 0 \}$ and $\overline{\CS}(k) = \{ \CS \in \Tilde{\CS} \; \big| \; k \not\in \CS \}$, 
for a generic network with cache placement matrix $\BV$ we have
\begin{equation}
    \CL_{BF}^{\BV} = \Bigg\{
    \frac
    {N_0 \alpha_{\CS_i^k} |\Bh_k^T \Bv_{\CS_i^k}|^2}
    {\sum\limits_{\CS_j^k \in \overline{\CS}(k)} \alpha_{\CS_j^k} |\Bh_k^T \Bv_{\CS_j^k}|^2 + N_0} 
    ; \;
    \begin{aligned}
        \forall k &\in [K] \\ \forall \CS_i^k &\in \CS(k)
    \end{aligned}
    \Bigg\} .
\end{equation}
%where $\CS(k) = \{ \CS \in \Tilde{\CS} \; \big| \;  k \in \CS ; \; \exists p \in \Phi(\CS): \BV[p,k] = 0 \}$ and $\overline{\CS}(k) = \{ \CS \in \Tilde{\CS} \; \big| \; k \not\in \CS \}$.

\subsection{Generic Networks}
Consider a generic network with parameters $K,L,t$, in which $K = t+L$ and cache placement is performed according to placement matrix $\BV$, with dimensions $P \times K$. Then, the total delivery time for EP, PL and BF beamformer structures at low-SNR can be approximately calculated as
\begin{equation}
\label{eq:generic_delivery_time}
    \begin{aligned}
        T_{EP}^{\BV} &= \frac{1}{P} \frac{N_0}{Po} \frac{n(\BV)}{\min \CL_{PE}^{\BV}} \; , \\
        T_{PL}^{\BV} &= \frac{1}{P} \frac{N_0}{Po} \frac{1}{\max\limits_{\alpha_{\CS}} \min \CL_{PL}^{\BV}} \; , \\
        T_{BF}^{\BV} &= \frac{1}{P} \frac{N_0}{Po} \frac{1}{\max\limits_{\alpha_{\CS},\Bv_{\CS}} \min \CL_{BF}^{\BV}} \; ,
    \end{aligned}
\end{equation}
where $n(\BV)$ is the number of $\CS$ sets for which $\Phi(\CS) \neq \varnothing$; and for any $\gamma \in \{ \mathrm{EP}, \mathrm{PL}, \mathrm{BF} \}$, $|\CL_{\gamma}^{\BV}| = K \times m(\BV) = PL$.

\section{Numerical Results and Discussion}
%Using delivery time expressions for generic setups in \eqref{eq:generic_delivery_time}, we investigate how the selection of subpacketization $P$ and beamformer structure $\gamma$ affects the system performance.
%We first start by $P$; and then explore the effect of $\gamma \in \mathrm{EP},\mathrm{PL},\mathrm{BF}$.
%
According to \eqref{eq:generic_delivery_time}, 
%it can be verified that the 
delivery time $T_{\gamma}^{\BV}$ is proportional to $\frac{1}{P}$; i.e increasing $P$ decreases $T_{\gamma}^{\BV}$. In fact, increasing $P$ enables %the requested 
data to be delivered in smaller chunks in parallel (over the MAC channel); which is characterized as \textit{efficiency index} in \cite{salehi2019subpacketization}. However, increasing $P$ also increases $|\CL_{\gamma}^{\BV}|$, 
%(the number of terms in $\CL_{\gamma}^{\BV}$), 
resulting in $\min \CL_{\gamma}^{\BV}$ to become smaller and $T_{\gamma}^{\BV}$ to be increased.

On the other hand, beamformer structure $\gamma$ only affects $T_{\gamma}^{\BV}$ through $\CL_{\gamma}^{\BV}$. For $\gamma = \mathrm{EP}$, the value of $\CL_{EP}^{\BV}$ is deterministic. However, if $\gamma \neq \mathrm{PL}$, $\min \CL_{\gamma}^{\BV}$ is a function of $\alpha_{\CS}$ (and $\Bv_{\CS}$, if $\gamma = \mathrm{BF}$). So $T_{\gamma}^{\BV}$ is calculated through an optimization problem, with $n(\BV)$ variables $\alpha_{\CS}$ and one constraint $\sum_{\CS} \alpha_{\CS} = 1$ for $\gamma = \mathrm{PL}$; and $(L+1)n(\BV)$ variables ($\Bv_{\CS} \in \mathbb{C}^L$) and $1+n(\BV)$ total constraints ($\|\Bv_{\CS}\| = 1$), for $\gamma = \mathrm{BF}$.

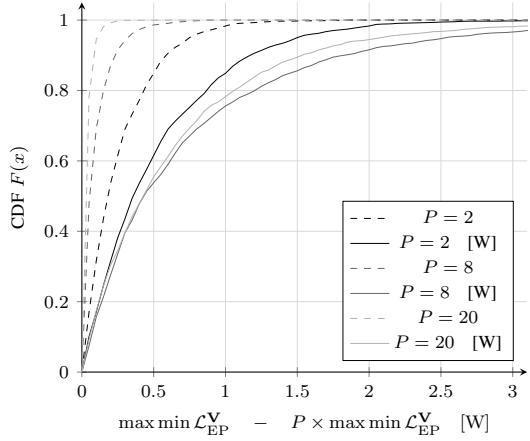
\begin{figure}[t]
    \centering
    \resizebox{0.8\columnwidth}{!}{%

    \begin{tikzpicture}

    \begin{axis}
    [
    % put axis lines at left and bottom
    axis lines = left,
    % control axis labels
    xlabel = \smaller {$\max \min \CL_{\mathrm{EP}}^{\BV} \quad - \quad P \times \max \min \CL_{\mathrm{EP}}^{\BV} \quad \mathrm{[W]}$},
    ylabel = \smaller {CDF $F(x)$},
    ylabel near ticks,
    % define axis ranges
    %xtick={0,1,...,10},
    %ytick={-8,-7,...,10},
    %xmin=0,
    xmax=3.1,
    ymax=1.05,
    % control legend position
    legend pos = south east,
    % control size of tick marks (10,20,30,etc)
    ticklabel style={font=\smaller},
    % control major grids
    grid=both,
    major grid style={line width=.2pt,draw=gray!30},
    % control minor grids
    %grid style={line width=.1pt, draw=gray!10},
    %minor tick num=5,
    ]
    
    \addplot
    [line width=0.15mm,dashed,black]
    table[y=N2,x=x]
    {Figures/Data/K6EP.txt};
    \addlegendentry{\smaller $P = 2$}

    \addplot
    [line width=0.15mm,black]
    table[y=2,x=x]
    {Figures/Data/K6EP.txt};
    \addlegendentry{\smaller $P = 2$ \; [W]}

    \addplot
    [line width=0.15mm,dashed,black!60]
    table[y=N8_1,x=x]
    {Figures/Data/K6EP.txt};
    \addlegendentry{\smaller $P = 8$}
    
    \addplot
    [line width=0.15mm,black!60]
    table[y=8_1,x=x]
    {Figures/Data/K6EP.txt};
    \addlegendentry{\smaller $P = 8$ \; [W]}
    
    \addplot
    [line width=0.15mm,dashed,black!30]
    table[y=N20,x=x]
    {Figures/Data/K6EP.txt};
    \addlegendentry{\smaller $P = 20$}
    
    \addplot
    [line width=0.15mm,black!30]
    table[y=20,x=x]
    {Figures/Data/K6EP.txt};
    \addlegendentry{\smaller $P = 20$ \; [W]}
    
    \end{axis}

    \end{tikzpicture}
    }

    \caption{Empirical CDF, $\gamma = \mathrm{EP}$}
    \label{fig:ep_plots}
\end{figure}

%the variables $\alpha_{\CS}$ are adjusted through an optimization problem, to maximize $\min \CL_{PL}^{\BV}$ (and decrease the delivery time). So for PL beamformer policy, we have $n(\BV)$ variables $\alpha_{\CS}$ and one extra constraint $\sum_{\CS} \alpha_{\CS} = 1$; which enables the delivery time to be smaller than EP policy. On the other hand, if $\gamma = \mathrm{BF}$, extra variables $\Bv_{\CS}$ can also be used for the optimization, resulting in even smaller delivery time. In fact, for BF policy, we have $(L+1)n(\BV)$ total variables (each $\Bv_{\CS}$ has $L$ elements), and $1+n(\BV)$ total constraints (for each $\Bv_{\CS}$, we should have $\|\Bv_{\CS}\| = 1$).
%
To compare the system performance for various $P$ and $\gamma$ selections, we use the cumulative distribution function (CDF) of $\max \min \CL_{\gamma}^{\BV}$, denoted by $F(\CL_{\gamma}^{\BV})$ for simplicity. As the terms inside $\CL_{\gamma}^{\BV}$ are correlated to each other, it is difficult to find a closed-form expression for $F(\CL_{\gamma}^{\BV})$; and so we proceed with a numerical approach.
%
%Unfortunately, this cannot be readily done as a closed-form expression; as the elements of $\CL_{\gamma}^{\BV}$ are correlated to each other, even if the channel vectors are independent. Here we take a numerical approach for finding the distribution of $\min \CL_{\gamma}^{\BV}$, for different $P$ and $\gamma$ selections.
%
In Figures \ref{fig:ep_plots}-\ref{fig:bf_plots} we have plotted the empirical results for $F(\CL_{\gamma}^{\BV})$ and $P \times F(\CL_{\gamma}^{\BV})$, for $\gamma \in \{ \mathrm{EP}, \mathrm{PL}, \mathrm{BF} \}$ and $P \in \{ 2,8,20 \}$. Network parameters are set to $K=6$, $t=3$, $L=3$. 
%. Note that the plot scaling is different in Figure \ref{fig:bf_plots} 
%(for $\gamma \in \{\mathrm{PL},\mathrm{BF}\}$ the results are multiplied by $n(\BV)$ also).

According to the figures, regardless of $\gamma$, the value of $F(\CL_{\gamma}^{\BV})$ decreases as we increase $P$. This is due to the fact that increasing $P$ increases $|\CL_{\gamma}^{\BV}|$; i.e. the minimum is taken over a larger number of random variables, resulting in higher probability for a smaller result.
However, comparing $P \times F(\CL_{\gamma}^{\BV})$ reveals that the decrement in $F(\CL_{\gamma}^{\BV})$ is dependent on $\gamma$ indeed. In fact, for $\gamma = \mathrm{EP}$, the decrement in $F(\CL_{\gamma}^{\BV})$ from $P=8$ to $P=20$ is so large that even multiplication by $P$ is not enough for its compensation. However, for $\gamma = \mathrm{PL}$ the decrement is almost compensated after the multiplication; and for $\gamma = \mathrm{BF}$ the compensation is so large that $F(\CL_{\gamma}^{\BV})$ is improved by increasing $P$. This is in line with the rate behavior with respect to $P$ and $\gamma$, as reviewed in Section \ref{sec:motivation}\footnote{As mentioned in Section \ref{sec:analysis}, the low-SNR approximation has a more positive effect on $R$ as $P$ becomes larger. So doing the analysis without the approximation causes more destruction in rate for larger $P$.}.

\begin{figure}[t]
    \centering
    \resizebox{0.8\columnwidth}{!}
    {

    \begin{tikzpicture}

    \begin{axis}
    [
    %width = 0.90\columnwidth,
    %height = 0.65\columnwidth,
    % put axis lines at left and bottom
    axis lines = left,
    % control axis labels
    xlabel = \smaller {$\max \min \CL_{\mathrm{PL}}^{\BV} \quad - \quad P \times \max \min \CL_{\mathrm{PL}}^{\BV} \quad \mathrm{[W]}$},
    ylabel = \smaller {CDF $F(x)$},
    ylabel near ticks,
    % define axis ranges
    %xtick={0,1,...,10},
    %ytick={-8,-7,...,10},
    %xmin=0,
    xmax=3.1,
    ymax=1.05,
    % control legend position
    legend pos = south east,
    % control size of tick marks (10,20,30,etc)
    ticklabel style={font=\smaller},
    % control major grids
    grid=both,
    major grid style={line width=.2pt,draw=gray!30},
    % control minor grids
    %grid style={line width=.1pt, draw=gray!10},
    %minor tick num=5,
    ]
    
    \addplot
    [line width=0.15mm,dashed,black]
    table[y=N2,x=x]
    {Figures/Data/K6PL.txt};
    \addlegendentry{\smaller $P = 2$}

    \addplot
    [line width=0.15mm,black]
    table[y=2,x=x]
    {Figures/Data/K6PL.txt};
    \addlegendentry{\smaller $P = 2$ \; [W]}

    \addplot
    [line width=0.15mm,dashed,black!60]
    table[y=N8_1,x=x]
    {Figures/Data/K6PL.txt};
    \addlegendentry{\smaller $P = 8$}
    
    \addplot
    [line width=0.15mm,black!60]
    table[y=8_1,x=x]
    {Figures/Data/K6PL.txt};
    \addlegendentry{\smaller $P = 8$ \; [W]}
    
    \addplot
    [line width=0.15mm,dashed,black!30]
    table[y=N20,x=x]
    {Figures/Data/K6PL.txt};
    \addlegendentry{\smaller $P = 20$}
    
    \addplot
    [line width=0.15mm,black!30]
    table[y=20,x=x]
    {Figures/Data/K6PL.txt};
    \addlegendentry{\smaller $P = 20$ \; [W]}
    
    \end{axis}

    \end{tikzpicture}
    }

    \caption{Empirical CDF, $\gamma = \mathrm{PL}$}
    \label{fig:pl_plots}
\end{figure}

\begin{figure}[t]
    \centering
    \resizebox{0.8\columnwidth}{!}{%

    \begin{tikzpicture}

    \begin{axis}
    [
    % put axis lines at left and bottom
    axis lines = left,
    % control axis labels
    xlabel = \smaller {$\max \min \CL_{\mathrm{BF}}^{\BV} \quad - \quad P \times \max \min \CL_{\mathrm{BF}}^{\BV} \quad \mathrm{[W]}$},
    ylabel = \smaller {CDF $F(x)$},
    ylabel near ticks,
    % define axis ranges
    %xtick={0,1,...,10},
    %ytick={-8,-7,...,10},
    %xmin=0,
    xmax=12.4,
    ymax=1.05,
    % control legend position
    legend pos = south east,
    % control size of tick marks (10,20,30,etc)
    ticklabel style={font=\smaller},
    % control major grids
    grid=both,
    major grid style={line width=.2pt,draw=gray!30},
    % control minor grids
    %grid style={line width=.1pt, draw=gray!10},
    %minor tick num=5,
    ]
    
    \addplot
    [line width=0.15mm,dashed,black]
    table[y=N2,x=x]
    {Figures/Data/K6BF.txt};
    \addlegendentry{\smaller $P = 2$}

    \addplot
    [line width=0.15mm,black]
    table[y=2,x=x]
    {Figures/Data/K6BF.txt};
    \addlegendentry{\smaller $P = 2$ \; [W]}

    \addplot
    [line width=0.15mm,dashed,black!60]
    table[y=N8_1,x=x]
    {Figures/Data/K6BF.txt};
    \addlegendentry{\smaller $P = 8$}
    
    \addplot
    [line width=0.15mm,black!60]
    table[y=8_1,x=x]
    {Figures/Data/K6BF.txt};
    \addlegendentry{\smaller $P = 8$ \; [W]}
    
    \addplot
    [line width=0.15mm,dashed,black!30]
    table[y=N20,x=x]
    {Figures/Data/K6BF.txt};
    \addlegendentry{\smaller $P = 20$}
    
    \addplot
    [line width=0.15mm,black!30]
    table[y=20,x=x]
    {Figures/Data/K6BF.txt};
    \addlegendentry{\smaller $P = 20$ \; [W]}
    
    \end{axis}

    \end{tikzpicture}
    }

    \caption{Empirical CDF, $\gamma = \mathrm{BF}$}
    \label{fig:bf_plots}
\end{figure}

\section{Conclusion and Future Work}
We studied the joint effect of subpacketization $P$ and beamformer structure $\gamma$ on the rate performance of multi-antenna Coded Caching (CC) setups. Using a low-SNR approximation, we provided simple closed-form rate expressions, and used numerical simulations for performance comparison of various schemes. The results indicate that $P$ and $\gamma$ jointly affect the rate; i.e. based on $\gamma$, $P$ might improve or deteriorate the achievable rate.
The results are limited to a specific class of networks with $K=t+L$. Removing this constraint and taking a more theoretical approach (compared to the numerical one in this paper), are parts of the ongoing research.

\bibliographystyle{IEEEtran}
\bibliography{References}

\begin{thebibliography}{10}
\providecommand{\url}[1]{#1}
\csname url@samestyle\endcsname
\providecommand{\newblock}{\relax}
\providecommand{\bibinfo}[2]{#2}
\providecommand{\BIBentrySTDinterwordspacing}{\spaceskip=0pt\relax}
\providecommand{\BIBentryALTinterwordstretchfactor}{4}
\providecommand{\BIBentryALTinterwordspacing}{\spaceskip=\fontdimen2\font plus
\BIBentryALTinterwordstretchfactor\fontdimen3\font minus
  \fontdimen4\font\relax}
\providecommand{\BIBforeignlanguage}[2]{{%
\expandafter\ifx\csname l@#1\endcsname\relax
\typeout{** WARNING: IEEEtran.bst: No hyphenation pattern has been}%
\typeout{** loaded for the language `#1'. Using the pattern for}%
\typeout{** the default language instead.}%
\else
\language=\csname l@#1\endcsname
\fi
#2}}
\providecommand{\BIBdecl}{\relax}
\BIBdecl

\bibitem{cisco2018cisco}
V.~N.~I. Cisco, ``{Cisco visual networking index: Forecast and trends,
  2017--2022},'' \emph{White Paper}, vol.~1, 2018.

\bibitem{6Genesis2019KeyIntelligence}
{6Genesis}, ``{Key Drivers and Research Challenges for 6G Ubiquitous Wireless
  Intelligence},'' \emph{White Paper}, vol.~1, 2019.

\bibitem{maddah2014fundamental}
M.~A. Maddah-Ali and U.~Niesen, ``{Fundamental limits of caching},'' \emph{IEEE
  Transactions on Information Theory}, vol.~60, no.~5, pp. 2856--2867, 2014.

\bibitem{shariatpanahi2018physical}
S.~P. Shariatpanahi, G.~Caire, and B.~H. Khalaj, ``{Physical-layer schemes for
  wireless coded caching},'' \emph{IEEE Transactions on Information Theory},
  vol.~65, no.~5, pp. 2792--2807, 2018.

\bibitem{karamchandani2016hierarchical}
N.~Karamchandani, U.~Niesen, M.~A. Maddah-Ali, and S.~N. Diggavi,
  ``{Hierarchical coded caching},'' \emph{IEEE Transactions on Information
  Theory}, vol.~62, no.~6, pp. 3212--3229, 2016.

\bibitem{pedarsani2016online}
R.~Pedarsani, M.~A. Maddah-Ali, and U.~Niesen, ``{Online coded caching},''
  \emph{IEEE/ACM Transactions on Networking (TON)}, vol.~24, no.~2, pp.
  836--845, 2016.

\bibitem{lampiris2018adding}
E.~Lampiris and P.~Elia, ``{Adding transmitters dramatically boosts
  coded-caching gains for finite file sizes},'' \emph{IEEE Journal on Selected
  Areas in Communications}, vol.~36, no.~6, pp. 1176--1188, 2018.

\bibitem{salehi2019coded}
M.~Salehi, A.~T{\"{o}}lli, and S.~P. Shariatpanahi, ``{Coded Caching with
  Linear Subpacketization is Possible in Multi-Antenna Communications},''
  \emph{arXiv preprint arXiv:1910.10384}, 2019.

\bibitem{tolli2018multicast}
A.~T{\"{o}}lli, S.~P. Shariatpanahi, J.~Kaleva, and B.~Khalaj, ``{Multicast
  beamformer design for coded caching},'' in \emph{2018 IEEE International
  Symposium on Information Theory (ISIT)}.\hskip 1em plus 0.5em minus
  0.4em\relax IEEE, 2018, pp. 1914--1918.

\bibitem{salehi2019subpacketization}
M.~Salehi, A.~T{\"{o}}lli, S.~P. Shariatpanahi, and J.~Kaleva,
  ``{Subpacketization-Rate Trade-off in Multi-Antenna Coded Caching},'' in
  \emph{2019 IEEE Global Communications Conference (GLOBECOM)}.\hskip 1em plus
  0.5em minus 0.4em\relax IEEE, 2019, pp. 1--6.

\bibitem{shariatpanahi2016multi}
S.~P. Shariatpanahi, S.~A. Motahari, and B.~H. Khalaj, ``{Multi-server coded
  caching},'' \emph{IEEE Transactions on Information Theory}, vol.~62, no.~12,
  pp. 7253--7271, 2016.

\bibitem{tolli2017multi}
A.~T{\"{o}}lli, S.~P. Shariatpanahi, J.~Kaleva, and B.~Khalaj, ``{Multi-antenna
  interference management for coded caching},'' \emph{arXiv preprint
  arXiv:1711.03364}, 2017.

\end{thebibliography}

\end{document}